# Nova Scorpius 1437 A.D. is now a dwarf nova, age-dated by its proper motion

M. M. Shara[1,2], K. Iłkiewicz[3], J. Mikołajewska[3], A. Pagnotta[1], M. Bode[4], L. A. Crause[5], K. Drozd[3], J. Faherty[1], I. Fuentes-Morales[6], J. E. Grindlay[7], A. F. J. Moffat[8], L. Schmidtobreick[9], F. R. Stephenson[10], C. Tappert[6], D. Zurek[1]

[1]Department of Astrophysics, American Museum of Natural History, CPW & 79th street, New York, NY 10024, USA

[2]Institute of Astronomy, The Observatories, Madingley Road, Cambridge CB3 0HA, UK

[3]N. Copernicus Astronomical Center, Polish Academy of Sciences, Bartycka 18, PL 00--716 Warsaw, Poland

[4]Astrophysics Research Institute, Liverpool John Moores University, IC2 Liverpool Science Park, Liverpool, L3 5RF, UK

[5]South African Astronomical Observatory, PO Box 9, Observatory 7935, Cape Town, South Africa

[6]Instituto de Física y Astronomía, Universidad de Valparaíso, Avda. Gran Bretaña 1111, 2360102 Valparaíso, Chile

[7]Harvard-Smithsonian Center for Astrophysics, The Institute for Theory and Computation, 60 Garden Street, Cambridge, MA 02138, USA

[8]Département de Physique and Centre de recherche en astrophysique du Québec (CRAQ), Université de Montréal, CP 6128 Succ. C-V, Montréal, QC H3C 3J7, Canada

[9]European Southern Observatory, Alonso de Cordova 3107, 7630355 Santiago, Chile

[10]Department of Physics, Durham University, South Road, Durham DH1 3LE, UK

**Classical novae are stellar outbursts powered by thermonuclear runaways in the hydrogen-rich envelopes of white-dwarf stars[1,2], accreted from close-binary red dwarf[3] or red giant[4] companions. Achieving peak luminosities up to 1 Million times that of the Sun[5], all classical novae are recurrent, on timescales from months[6] to millennia[7]. During the century before and after a nova eruption, the binaries that give rise to classical novae – novalike binaries – almost always exhibit high rates of mass transfer to their white dwarfs[8]. Binaries with stellar masses and periods indistinguishable from those of novalikes[9], but with much lower mass transfer rates[10], exhibit dwarf nova outbursts when accretion-disk instabilities[11] drop matter onto the systems' white dwarfs. The simultaneous existence of novalike binaries and dwarf novae, cataclysmic variable binaries that are identical but for widely varying accretion rates, has been a longstanding puzzle[9]. One scenario[12] predicted that novalike binaries undergo a transition to become dwarf novae several centuries[13] after classical nova eruptions. The few known examples[14,15,16,17] of this post-nova-to-dwarf nova metamorphosis lack robust and independent determinations of the date of the associated classical nova[18,19], and hence the time required to undergo the high-to-low mass transfer transition. Here we report the recovery of the binary underlying the classical nova of 11 March 1437 A.D.[20,21] whose age is independently confirmed by proper motion-dating, and show that in the 20th century it exhibits dwarf nova eruptions. The four oldest recovered classical novae are now all dwarf novae. Taken together they strongly suggest that mass transfer rates decrease by an order of magnitude or more in the centuries after a classical nova event, greatly slowing the evolution, and lengthening the lifetimes of these explosive binary stars.**

**Subject terms: Stars Novae**

The long-term evolution of cataclysmic variable binaries is uncertain. Some may produce Type Ia supernovae[22], while others likely consume their red dwarf secondaries[23]. The most important parameter controlling the evolution and lifetime of a cataclysmic variable is dM/dt, the system's average rate of mass-transfer from the companion donor star to its white dwarf. Remarkably, at any given binary orbital period, a wide range of dM/dt values is observed[9].

Almost all cataclysmic variables are novalike binaries, displaying high values of dM/dt, in the century before and after classical nova eruptions[8]. The hibernation scenario[12] of cataclysmic variable evolution postulated that dM/dt decreases, after classical nova eruptions, on a timescale of several centuries as irradiation of the red dwarf, from its cooling white dwarf companion, decreases[13]. The scenario thus predicted that novalike binaries and dwarf novae are really the same systems, seen at different times after their most recent classical nova eruption, and that all classical novae eventually display dwarf nova behavior. There is a handful of classical novae that display dwarf nova outbursts a few decades after eruption[23], but the vast majority do not. The oldest classical nova whose eruption was observed is WY Sge (Nova 1783 A.D.)[24]; it remains a high dM/dt novalike binary[25] 234 years after its eruption. Three dwarf novae in old nova shells have been found[14,15,17], but the dates of the last classical nova eruptions for all three are uncertain. Thus the observed transition time from classical nova and novalike binary to dwarf nova is almost always longer than two centuries, but is otherwise unconstrained.

One of the best-located novae of antiquity, recorded by Korean royal astronomers, erupted on 11 March 1437 A.D[26]. It lay within the asterism Wei (the tail of the modern constellation Scorpius), within half a chi (roughly 1 degree) of one of the two stars ζ Sco or η Sco (See

Methods). It was seen for 14 days before vanishing, consistent with a fast-declining classical nova while ruling out a supernova.

A search of a 1985 U and B-band Anglo-Australian 1.2 meter Schmidt telescope plate-pair centered on Scorpius yielded dozens of UV-bright objects, but only one with a clear, shell-like structure. The shell is marginally visible on other sky survey plates, but clearly seen in a narrowband Hα image (See Figure 1). The central star of the nebula is *not* a cataclysmic variable. However, a star about 15" from the shell center is both the strong emission-line cataclysmic variable 2MASS J17012815-4306123 and the variable X-ray source IGR J17014-4306[27].

The measured proper motion of the 17th magnitude cataclysmic variable, listed in multiple catalogues, displays large scatter. This is because all such southern hemisphere catalogues to date are limited to short, 20-year baselines, and because the star is blended with the image of a nearby neighbour. We have located the star on a digitized Harvard DASCH[28] (Digital Access to a Sky Century @ Harvard) photographic plate A12425, taken in 1923 (see Figure 2 for details). The cataclysmic variable's position was measured relative to nearby stars both on that plate and on a sub-arcsecond CCD image taken in 2016 (see Methods). The 93-year baseline has enabled us to measure the cataclysmic variable's proper motion with far higher precision than previously possible:

$\mu_\alpha$ = -12.74 +/- 1.79 milliarcsec/yr, and $\mu_\delta$ = -27.72 +/- 1.21 milliarcsec/yr. ,

where the 1-σ errors are determined empirically from the proper motions of dozens of stars on both plates. This proper motion places the cataclysmic variable, in 1437 A.D. at:

RA (J2000) = 17:01:28.53 +/- 1.0",

DEC (J2000) = -43:05:56.7 +/- 0.7".

The center of the nova shell in 2016 is determined from its edges (See Figure 1 and Methods) to be at:

RA (J2000) = 17:01:28.55 +/- 1.43",

DEC (J2000) = -43:05:59.3 +/-0.4".

Unlike their underlying cataclysmic variables, nova shells decelerate by sweeping up interstellar matter, halving their speeds on a mean timescale of 75 years[29]. Assuming that the nova 1437 ejecta initially had the same proper motion that we measure for the cataclysmic variable, and that value has been halved every 75 years since, the shell center in 1437 must have been located +1.43" (east) in Right Ascension and +3.1" (north) in Declination of its 2016 position at:

RA (J2000) = 17:01:28.68 +/- 1.43",

DEC (J2000) = -43:05:56.2 +/-0.4".

The difference between the 1437 center of the nova shell and the proper motion-determined cataclysmic variable position in 1437 A.D. is 1.7". The 1$\sigma$ error ellipses of the two positions overlap, confirming that the cataclysmic variable is in fact the nova of 1437 A.D. and the source of the shell we observe today.

This independent, proper-motion based clock is unprecedented, enabling the first unambiguous identification and precise age dating of a pre-telescopic nova from Imperial Asian astronomical records. Fortuitously the cataclysmic variable is a deeply eclipsing system, enabling measurement of its orbital period $P_{orb}$=0.5340055 (+/-5e-7) days, and detailed characterization of its stellar components (see Methods).

A DASCH light curve of the cataclysmic variable from 1919 to 1951 is shown in Figure 3. Three dwarf nova eruptions (in 1934, 1935 and 1942) are clearly seen, wherein the cataclysmic variable brightened by 2 - 4 magnitudes. In Figure 4 we show images before, during and after one of those dwarf nova eruptions. The classical nova of 1437 A.D, seen 497 years after eruption, has become a dwarf nova. dM/dt in this precisely age-dated post-classical nova, and in all three other ancient post-classical novae (which have nova shells but lack accurately-known ages) are at least an order of magnitude less than those displayed by most novalike binaries one to two centuries after they erupt. This is the strongest evidence yet that novalike binaries and dwarf novae are indeed the same systems, seen at different times after classical nova explosions.

Future cataclysmic variable evolution simulations should incorporate this large decrease in dM/dt in the modeling of post-nova binaries. Many cataclysmic variables experience at least 15,000 years between classical nova eruptions[30] in a state of low dM/dt, before returning to the novalike binary, high dM/dt state leading up to their next classical nova eruption. The effect of much lower dM/dt is to greatly lengthen the time needed to accrete the critical-mass envelope which triggers the next classical nova eruption. This slows the evolution of cataclysmic variable binaries, and lengthens their lifetimes, as the reservoirs of fuel in red dwarfs will last much longer with their greatly decreased dM/dt.

**Methods** is linked to the online version of the paper at www.nature.com/nature.

**Acknowledgements**

JM, KI, KD and MMS gratefully acknowledge support by the Polish NCN grant DEC-2013/10/ M/ ST9/ 00086. AFJM is grateful to NSERC (Canada) and FQRNT (Quebec) for financial support. AP acknowledges support from the American Museum of Natural History's Kathryn W. Davis Postdoctoral Scholar program, which is supported in part by the New York State Education Department and by the National Science Foundation under grant numbers DRL-1119444 and DUE-1340006. MMS gratefully acknowledges the hospitality of the Institute of Astronomy at the University of Cambridge. Some of the observations reported in this paper were obtained with the Southern African Large Telescope (SALT) under programme 2016-1-SCI-044, and with the South African Astronomical Observatory's 1-meter telescope. Polish participation in SALT is funded by grant No. MNiSW DIR/WK/2016/07. We gratefully acknowledge the Harvard/CfA team that has made DASCH data available to the astronomical community.

The DASCH project at Harvard is partially supported from NSF grants AST-0407380, AST-0909073, and AST-1313370. IRAF is distributed by the National Optical Astronomy Observatories, which is operated by the Association of Universities for Research in Astronomy under a cooperative agreement with the National Science Foundation. This research has made use of data obtained from the Chandra Data Archive and software provided by the Chandra X-ray Center in the application packages CIAO, ChIPS, and Sherpa.

**Author Contributions:**

All authors shared in the ideas and writing of this paper. MMS, AFJM and MB carried out optical surveys for the nova based on F. R. Stephenson's predictions. MMS and AFJM located the nebula associated with the old nova. Broadband CCD imaging and data reduction of the candidate was carried out by LC, IF-M and KD. Narrowband imaging of the shell and reduction of those images was carried out by KI, who also produced the cataclysmic binary's X-ray light-curve and deduced its period. JG retrieved the 1923 digitized image of the nova. AP and JG produced the old nova's historical light curve. AP and KI measured the old nova's proper motion. MMS, KI and JM determined the orbital period, while IF-M, KI and JM determined the white dwarf spin period. JM determined the companion's spectral type, the system's mass function, and its distance, while JM and MMS found the limit on the shell's mass.

**Author Information**

Reprints and permission information is available at www.nature.com/reprints.

**Competing Financial Interests**

The authors declare no competing financial interests.

**Correspondence Author**

Correspondence should be addressed to MMS (mshara@amnh.org).

**Data Availability**

All relevant data are available from the corresponding author on reasonable request.

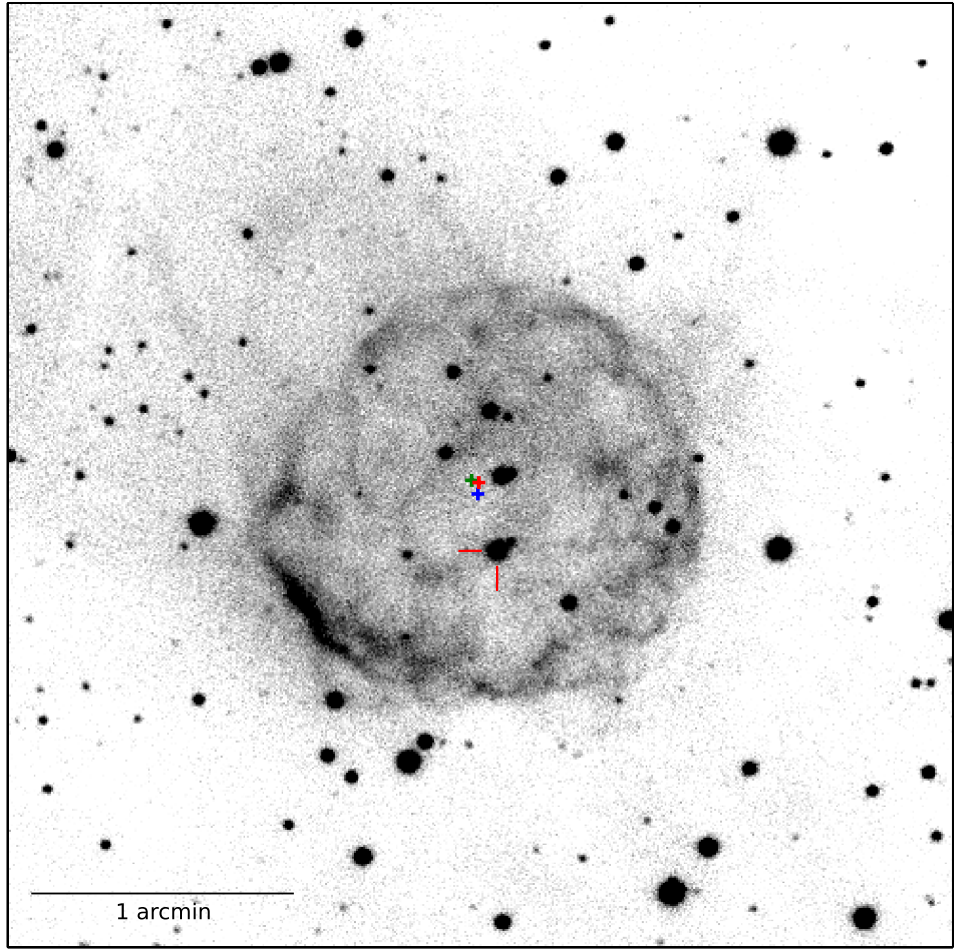

**FIGURE 1.** The recovered nova of 1437 A.D. and its ejected shell. The Hα image was taken with the *SWOPE* 1-meter telescope and its CCD camera in September 2016 with a total of 6000 seconds of exposure. North is up and east is to the left. The images were processed and combined with standard *PYRAF* and *IRAF* procedures. The cataclysmic variable is indicated with red tick marks. Its proper motion places it 7.4" East and 16.0" north of its current position, at the red ``+'' in 1437 A.D.. The position of the center of the shell in 2016 and its deduced position in 1437 A.D. (see main text) are indicated with blue and green plus signs, respectively. The 1437 A.D. positions of the shell center and of the cataclysmic variable agree to within 1.7", and their 1σ error ellipses overlap.

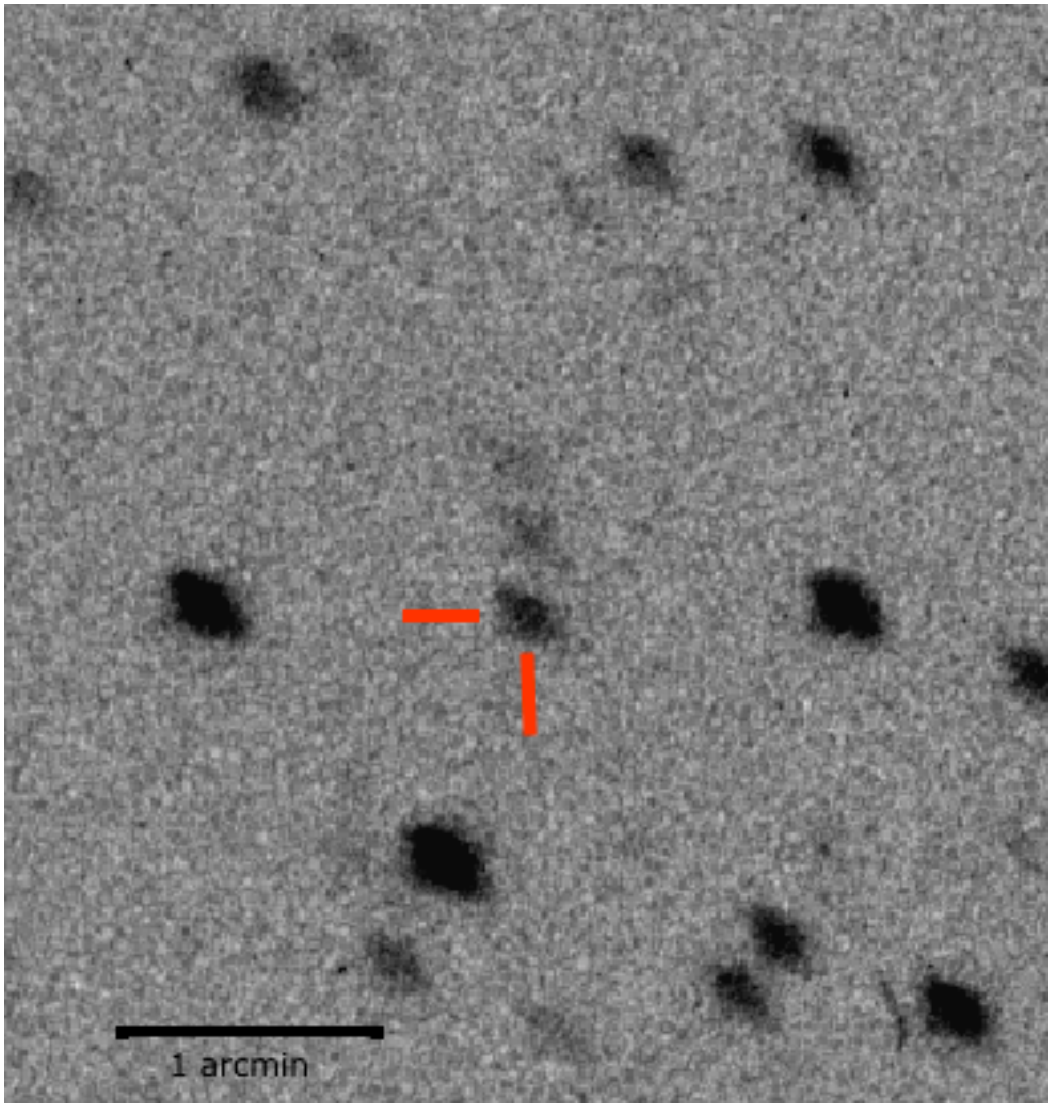

**Figure 2.** A section of Harvard plate A12425, a 300-minute exposure taken on 10 June 1923 using the 24" Bruce Doublet telescope at the Harvard Observatory station in Arequipa, Peru. North is up, East is to the left, and the cataclysmic variable is again indicated with red tick marks. This plate, like most at Harvard, is essentially equivalent to a B-band image.

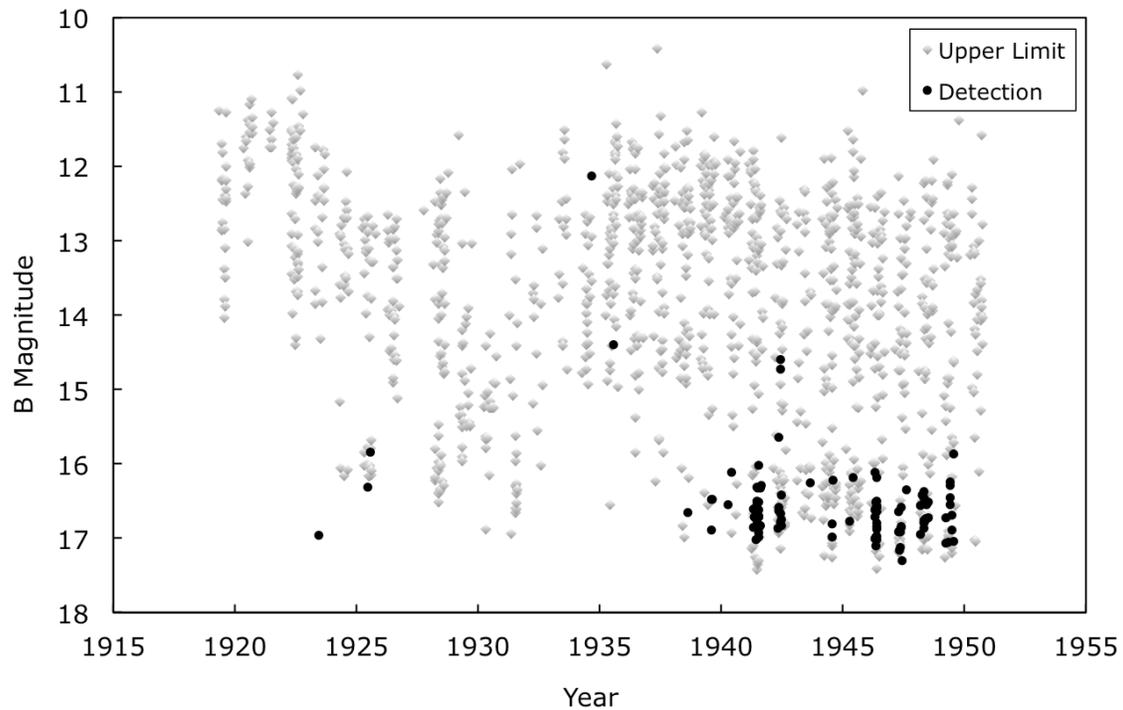

**Figure 3.** The 1919 – 1951 light curve of Nova Scorpius 1437 A.D. The gray symbols are upper limits, while the black dots are measured detections of the star. The star was first detected in quiescence near 17$^{th}$ magnitude in 1923, and near 16$^{th}$ magnitude in 1925. Dwarf nova outbursts in 1934 (reaching nearly 12$^{th}$ magnitude), 1935 and 1942 (see Figure 4) are evident.

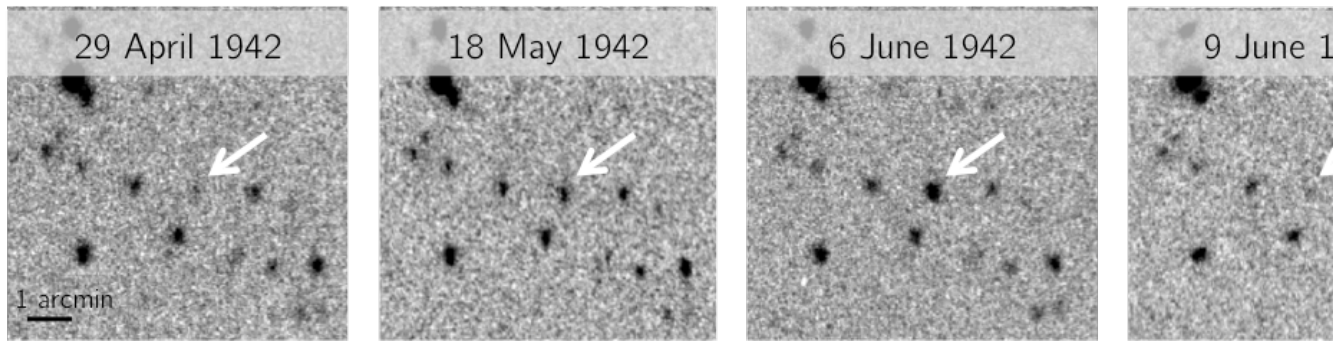

**Figure 4.** A 1942 dwarf nova outburst of nova Scorpius 1437 A.D. on a series of Harvard DASCH MF-series plates (see Methods). North is up, east is left. The dwarf nova is indicated with an arrow in each epoch.

**Methods**

1. The historical Nova Scorpius of 11 March, 1437 A.D.

2. The nova shell's center.

3. The cataclysmic variable orbital period and ephemeris

4. The white dwarf spin period

5. The white dwarf mass

6. The shell mass

## Methods

**1. The historical nova Scorpius of 11 March, 1437 A.D.**

The Sejong Sillok ("Veritable Records of [the reign of] King Sejong") is a detailed chronicle of the reign of King Sejong (who ruled Korea from A.D. 1418 to 1464), written in Classical Chinese. Chapter 76 of the Sillok notes:

"19th year of King Sejong, 2nd lunar month, day yichou [the 2nd day of the 60-day cycle], A meteor (liuxing) appeared... A solar halo... A guest star (kexing) began to be (shi) seen between the second and third stars of Wei. It was nearer to the third star, about half a chi ("half a foot") away. It lasted (jiu) for 14 days."

It was quite usual for East Asian astronomers to use "chi" - a linear unit - as an angular unit. Descriptions of the position of the ecliptic relative to the 28 Chinese constellations (or "mansions") through which the moon passes suggested that 1 chi, as used in China, was roughly 1.50 deg[31]. Early Chinese records of planetary conjunctions in which separations were quoted in chi suggest that the chi/degree ratio is very roughly one. However, values in the range 0.44 to 2.8[31] have been determined, so that half a chi in China was roughly in the range 0.22 – 1.4 degrees. There is no similar determination for the Korean chi.

The Sejong Sillok refers to the guest star as being between the second and third stars of Wei, while the date corresponds to 11 March 1437 A.D.. Wei, the 6th lunar lodge, contained nine stars in Scorpius[32]. Listed in clockwise order they are ε, μ, ζ, η, θ, ι, κ, λ and ν Sco. The numbering of the stars in Wei requires deduction as no star map from ancient Korea which lists star numbers in constellations exists. Korean astronomers adopted asterisms similar to those of their Chinese colleagues, so it is reasonable to look to Chinese star maps and lists for guidance. Two Chinese Yuan dynasty lists differ in their ordering:

(i) ε Sco (1), μ Sco (2), ζ Sco (3), η Sco (4), and:

(ii): μ Sco (1), ε Sco (2), ζ Sco (3), η Sco (4).

The reasonable deduction is that ζ Sco is the third star of Wei.

Since μ Sco is adjacent to ζ Sco, it is also reasonable to identify μ as the second star of Wei, as mentioned in the guest star text. The guest star would then be in the range 0.22 – 1.4 deg N of ζ Sco. Searches based on this predicted position have proven fruitless[20,21].

An alternative numbering of the stars of Wei is based on μ Sco, the determinant star of Wei, which fixed the boundary of the lunar lodge Right Ascension zone. Starting from μ Sco, and proceeding strictly clockwise around Wei, the second star in the text would be ζ Sco, while the third would be η Sco. If this numbering is correct, then one should find the old nova between ζ Sco and η Sco, i.e. to the East of ζ Sco. This is, in fact, where we find the cataclysmic variable and nova shell that are the subject of this letter. In light of the uncertainties noted above concerning the numbering of the stars in Wei in Korea, the size of the Chinese chi, and the even greater uncertainty of the size of that unit in Korea, the observed angular distances of 1.95 degrees and 1.55 degrees from the cataclysmic variable to η Sco and to ζ Sco, respectively, are in reasonable accord with the historical text.

## 2. The Nova Shell and its Centre

The nova shell was imaged with the 1.0m Swope telescope at Las Campanas observatory on 15 and 17 June 2016. The observations were carried out with an E2V CCD231-84 CCD camera with pixel size of 0.435 arcsec. The total exposure time was 6000 sec through a narrowband Hα filter. The images were reduced with standard IRAF procedures. The cataclysmic variable is not at the center of the nebula (Figure 1). The nova shell shows complex morphology, with a number of brighter knots, and strongest emission at the SE edge of the nebula. Moreover, there is a faint outer lobe of nebulosity to the NW of the nova shell. At the NE corner of the nebula there is a faint tail-like structure extending up to a few arcmin away from the nebula. Radial velocities will be required to distinguish shell material from diffuse interstellar gas emission, which is strong in the direction of Scorpius.

The outer edges of the H-alpha shell were used to determine the center of the shell[33]. An initial cut through the shell in the north/south direction was perpendicularly bisected and the center of the bisector was retained as the starting center position. A new cut through the starting center was made approximately 10 degrees clockwise from the starting cut, and then it was perpendicularly bisected, again retaining the center of the bisector as the next center position. This procedure was repeated twelve times until the center measurements converged to within a pixel of each other. The last five iterations--where the convergence was strongly evident--were averaged together to obtain the measured center, with the standard deviation of those five positions used as the 1-σ uncertainty on each measurement.

## 3. The Nova Orbital Period and Ephemeris

Photometric monitoring of the system was carried out with three telescopes. On 27, 29, 30 and 31 July 2016 we observed the old nova with the 2.5m Du Pont telescope at LCO with the SIT e2K-1 camera and a V filter. Exposure times were 30s on 31 July and 40s on the rest of the nights. On 15, 16, 20, 23, 24, 25 and 26 of August 2016 the object was observed with a Sutherland High-speed Optical Camera[34] on the 1.0m South African Astronomical Observatory telescope in the g' filter and with 4x4 binning. The exposure times were 15s on 20, 25 and 26 of August, 25s on 23 of August and 20s on other nights. On 12, 13, 20 and 21 September 2016 the nova was observed with the Swope telescope in B band. The exposure times were 180s. The data were processed with standard IRAF procedures. The magnitudes were transformed to a standard system using all of the stars in the field of view of each of the telescopes as standard stars. The reference magnitudes were taken from the American Association of Variable Star Observers' All-Sky Photometric Survey[35].

The cataclysmic variable light curve shows deep and short eclipses, and ellipsoidal variability (Fig. 1). We applied the Phase Dispersion Minimization[36] methodology, which is well suited to light curves with long gaps and relatively short eclipses; it determined the orbital period to be $P_{orb} = 0.534033d = 12h49m00.5s$. The orbital period was further refined to $P_{orb}=0.5340055d$ (+/-5e-7) using RI photometry from the Bochum Galactic Disk Survey[37]. We also fit a linear ephemeris to the measured times of eclipses, which gave:

HJD MIN=2457626.3643(+/-3e-4) + 0.534055(+/-5e-7)*E     (Eq. 1)

## 4. The White Dwarf Spin Period

In addition to the orbital-induced variability there is a periodic variability on a time scale of 30 minutes visible outside of the eclipses. We fitted and removed a low order polynomial from each dataset to remove orbital-induced variability. Discrete Fourier transforms performed on these altered light-curves with the Period04 program[38] resulted in an ephemeris

HJD MAX=2457625.059060(+/-6e-6)+0.0215175(+/-8e-7)*E    (Eq. 2)

for the maxima of this variability. Over the 11 days monitored by the Sutherland High-speed Optical Camera the period remained stable at 1859.112 seconds, demonstrating that this variability is not flickering, but instead due to the spin period of the white dwarf. This, in turn, suggests that the system is an intermediate polar (as does the presence of HeII emission – see below). The object shows two pulses per period, with different amplitudes (Fig. 2). The detailed analysis of the pulse profiles is beyond the scope of this paper, though we note that the pulse profiles varied slightly from night to night.

The nova was observed with the Chandra Advanced CCD Imaging Spectrometer (ACIS)[39], which covers the energy range 0.2 -10 keV. The observation was performed on 30 June 2015 and the total exposure time was 10.07 ks. We extracted the light-curve using the software programs CIAO version 4.8.1[40] and CALDB version 4.7.2. The adopted bin size of the light-curve was 120s. The object showed spin variability related to the spin period, similar to the one observed in the optical range (Fig. 2). We performed a discrete Fourier transform on the Chandra light-curve and calculated the errors using a Monte Carlo simulation with the Period04 program. The resultant period of 0.0218+/-0.0003d = 1859 +/- 26 sec is consistent with the period derived from the optical Sutherland High-speed Optical Camera data. The simultaneous presence of the same periodic variability in X-ray and optical bands, measured a year apart, confirms that this permanent period is indeed the spin period of the white dwarf.

## 5. The White Dwarf Mass

We obtained low resolution, long-slit spectra of the cataclysmic variable and its shell with the Robert Stobie Spectrograph (RSS)[41,42] and the 10m Southern African Large Telescope (SALT)[43,44]. We employed grating PG0900 and a slit with a projected width of 1.5" which resulted in a resolving power of R~1000. The observations were reduced using standard IRAF procedures and the SALT RSS science pipeline[45]. The brightest knot in the nova shell was observed on 24 September 2016. The cataclysmic variable was observed on 14 and 19 July 2016, and twice on 23 September 2016.

The spectra of the cataclysmic variable reveal broad (FW~2000-2700 km/s) emission lines of HI Balmer, HeII and HeI lines, and a wealth of absorption lines indicating the presence of an early K-type secondary (Fig. 3). The relative intensities of temperature-sensitive metal absorption lines as well as the simultaneous presence of G-band, MgH bands and a very weak TiO band at 6159 Å are consistent with a K3 star. This spectral classification is consistent with the broadband magnitudes and colors observed during the eclipses (B-V ~ 1.3-1.4) and 2MASS JHK magnitudes, and moderate reddening E(B-V) ~ 0.3-0.4.
The HI Balmer line ratios in the low-density nebula/shell surrounding the cataclysmic variable give similar E(B-V) ~ 0.3-0.6 (assuming case B recombination) and an absorption A(V) ~ 1. Since three of these spectra were fortuitously taken close to both spectroscopic quadratures, we can estimate the preliminary radial velocity amplitude of the secondary, $K_{sec}$ = 220 +/-27 km/s (assuming that the eclipses coincide with its inferior conjunction, and adopting the photometric ephemeris). The best fit to the four radial velocity data points results in a secondary velocity $K_{sec}$ = 260 +/-6 km/s, with the spectroscopic conjunction occurring 0.035 Porb after the eclipse (Fig. 4). The mass function is then:

$f(m) = 0.974^{+0.105}_{-0.055}$ $M_{sun}$ .

While the phase shift might be the result of small-sample statistics (only four radial velocity points), such a shift between spectroscopic conjunction and photometric eclipse is not unusual in magnetic cataclysmic variables. The eclipse in such cases is not that of the white dwarf but rather of the principal accretion spot, which can be displaced from the line connecting the two stars[46].

The mass function demonstrates that the white dwarf must be massive and the secondary must be evolved. Nova 1473 A.D. was rather fast, visible for only 14 days, so the white dwarf mass[7] is likely > 1.0 $M_{sun}$. For a white dwarf mass close to 1.0 $M_{sun}$, the companion mass will be in the range 0.3 - 0.76 $M_{sun}$; if the white dwarf mass is close to the Chandrasekhar limit then the secondary mass will be <~0.2 $M_{sun}$. Since the secondary is filling its Roche lobe we can estimate the system distance d. For a 1.0 $M_{sun}$ white dwarf, 650 <~ d <~980 pc, while for a 1.4 $M_{sun}$ white dwarf, d<~ 540 pc.

The cataclysmic variable emission lines have variable, complex profiles, but the changes do not seem to show orbital modulation. They may be varying on shorter timescales related to the 1859 sec white dwarf rotation period.

The spectrum of the brightest knot (SE of the cataclysmic variable) in the shell reveals many emission lines (see Fig. 2), including very strong (compared to Hα) [NII] and [SII]. [OIII] is relatively weak (fainter than Hβ) which is unusual given the presence of HeII 4686 emission. The [NII]/Hα, [SII]/Hα and [OIII]/Hβ flux line-ratios are intermediate between those of planetary nebulae, supernova remnants and HII regions, as seen in emission-line diagnostic diagrams[47]. The ratio of [SII] 6716:6731=1.45 indicates a very low electron density, $n_e < \sim 100$ cm$^{-3}$, whereas the [NII] lines ratio (6548+6583)/5755=100 implies an effective temperature $T_e = 9400$ Kelvins[48].

## 6. The Shell Mass

The H$\alpha$ +[NII] flux of the nova shell is 2.8 x $10^{-15}$ W/m$^2$, and our measurement of the H$\alpha$ /[NII] flux ratio is ~ 0.6 from our SALT spectra (Figure 2). The upper limit on density is $n_e$<~100 cm$^{-3}$ as noted above. This density and the distance derived above yield $n_e^2$ V = 2.35 x $10^{56}$ cm$^{-3}$ [d/500 pc]$^2$. Allowing for reddening of E(B-V) ~0.3 – 0.4 derived above, and hence A(V) ~1, finally yields an upper limit on the radiating hydrogen gas in the nova shell of $M_{shell}$ < 0.004 $M_{sun}$ [d/500 pc]$^2$.

The ejecta in nova shells decelerate to half their initial velocities by sweeping up interstellar matter and doubling their masses on timescales of about 75 years[29]. Even if the ejected shell only underwent two successive mass doublings (and it may have undergone 7 or 8 such doublings), a hard upper limit on the mass ejected in the nova eruption is $M_{ej}$ < $10^{-3}$ $M_{sun}$. This rules out any chance of the shell having a planetary nebula origin[49] because the masses of planetary nebula shells are typically 0.1 – 1.0 $M_{sun}$[50].

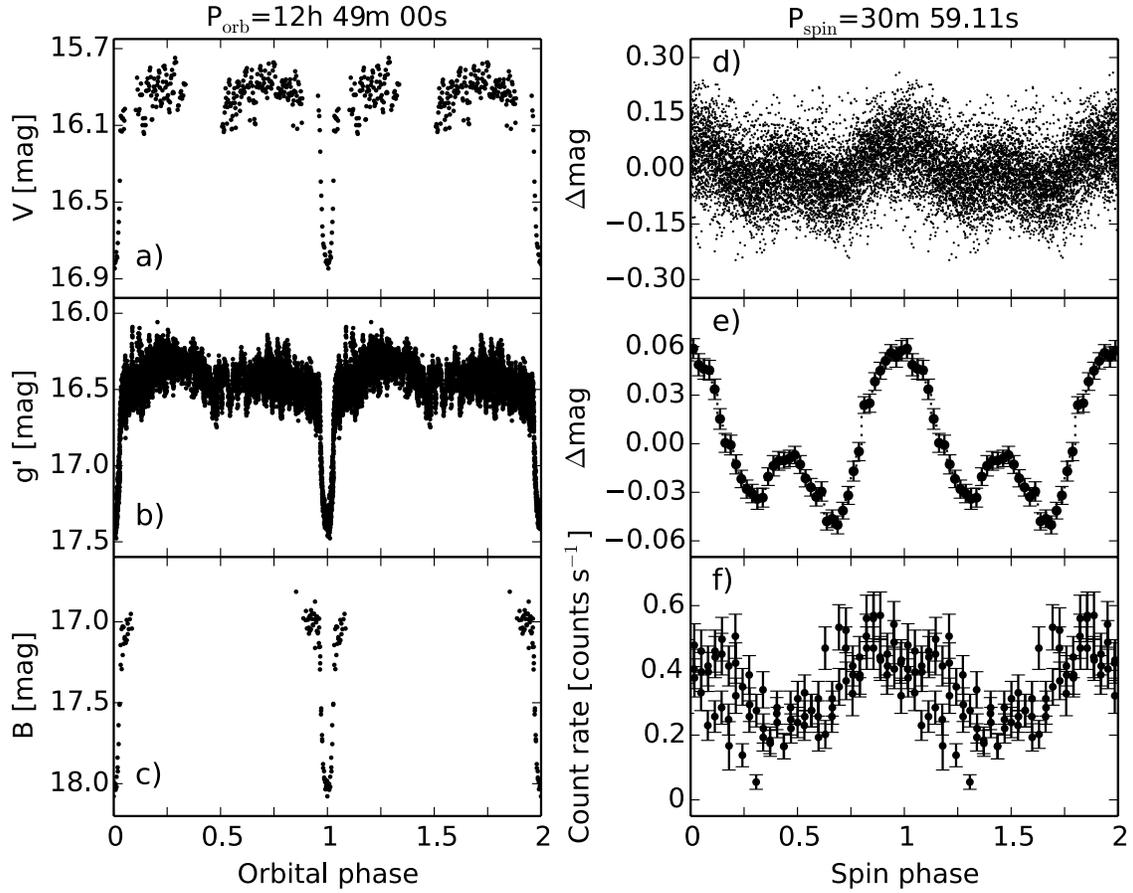

**Figure 1.** Panels a), b) and c): Phase plot of Nova 1437 photometry in V, g' and B bands respectively. The observations are phased with the orbital period, using the ephemeris of Eq. 1. Panel d): Phase plot of g' photometry, after subtracting all variability related to the orbital motion (see Methods section 4). The observations were phased with the spin period of the white dwarf using the ephemeris of Eq. 2. Panel e): Same as d), but with points binned with a bin size of $0.025 \times P_{spin}$. Panel f) Chandra observations in the 0.2-10 keV band phased with the spin period of the white dwarf using the ephemeris from Eq. 2.

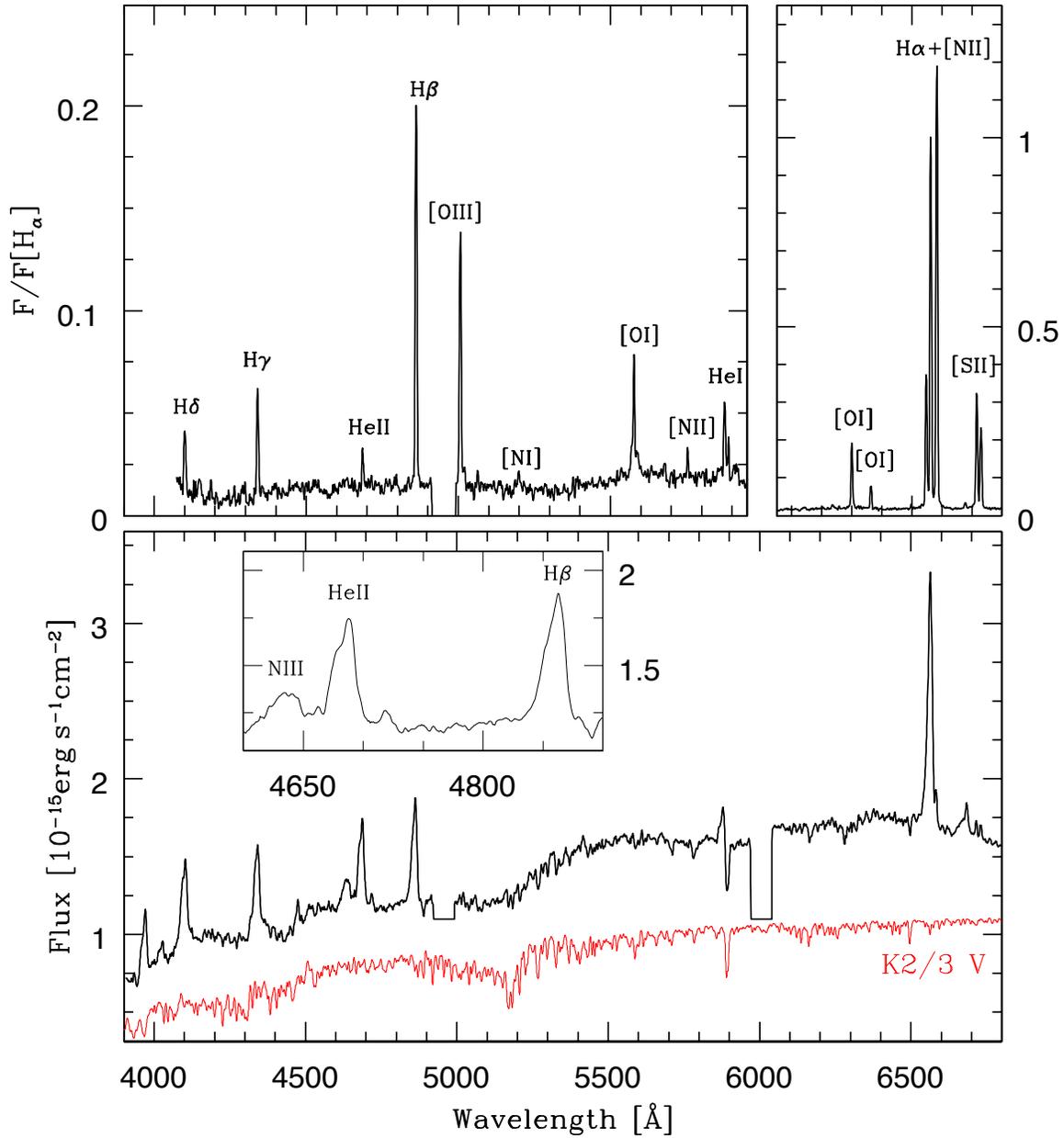

**Figure 2.** Upper panel: SALT spectrum of the brightest region on the nova shell (SE of the cataclysmic variable) with identifications of the main emission lines. Note the strong lines of [SII] 6716,6731 and [NII] 6548,6583.
Lower panel: SALT spectrum of the cataclysmic variable taken on 23 September 2016 with the overlaid synthetic spectrum of a K3 V star (with effective temperature Teff=4750 Kelvins, gravitational acceleration log g = 4.5, and solar composition) reddened with A(V)~1. The insert shows the emission profiles of Hβ as well as those of HeII and the Bowen's NIII blend.

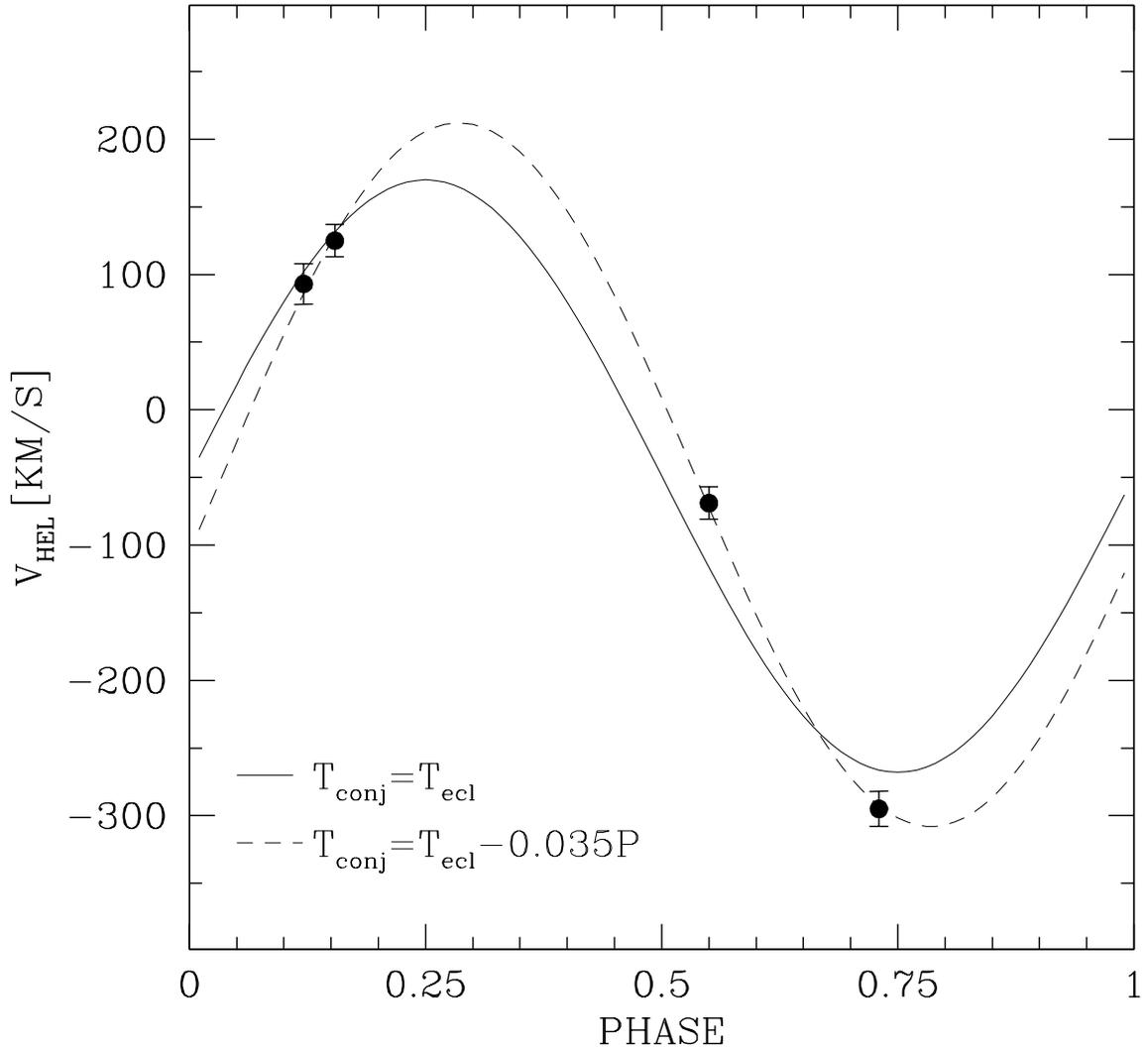

**Figure 3.** The Nova 1437 radial velocity curve. The solid curve corresponds to the secondary star's inferior conjunction occurring at mid-eclipse. The (better fitting) dashed curve corresponds to inferior conjunction occurring 0.035P after the eclipse. See text for details.